\documentclass[a4paper]{jpconf}
\usepackage{graphicx}
\usepackage{epstopdf}
\usepackage{graphicx,epsf,amsmath,amsfonts,amssymb,amsbsy,enumerate}

\newcommand{\ds}{\displaystyle}
\newcommand{\vev}[1]{\langle#1\rangle}
\newcommand{\emat}{\end{array} \right )}
\newcommand{\vect}{\left ( \begin{array}{c}}
\newcommand{\evect}{\end{array} \right )}

\begin{document}
\title{QCD phase diagram with chiral imbalance in %the framework of 
NJL model: duality and lattice QCD results}

\author{T. G. Khunjua}
\address{University of Georgia, Tbilisi, Georgia; Faculty of Physics, Moscow State University,
119991, Moscow, Russia}
\author{ K. G. Klimenko}
\address{Logunov Institute for High Energy Physics,
NRC "Kurchatov Institute", 142281, Protvino, Moscow Region, Russia}

\author{ Roman N. Zhokhov}

\address{ %State Research Center of Russian Federation -- Institute for High Energy Physics, NRC "Kurchatov Institute", 142281, Protvino, Moscow Region, Russia,  IZMIRAN, Moscow, Russia;
-Logunov Institute for High Energy Physics,
NRC "Kurchatov Institute", 142281, Protvino, Moscow Region, Russia;

-Pushkov Institute of Terrestrial Magnetism, Ionosphere and Radiowave Propagation (IZMIRAN),
108840 Troitsk, Moscow, Russia}

\ead{zhokhovr@gmail.com}

\begin{abstract}
 In addition to temperature and baryon chemical potential there are other parameters that matter in the real quark matter. One of them is isospin asymmetry which does exist in nature, for example, in the compact stars and in heavy ion collisions. The chiral imbalance, the difference between left- and right-handed quarks, is another phenomenon that could occur in quark matter. We have shown that lattice QCD results support the existence of the approximate duality in real QCD. The rise of pseudo-critical temperature with increase of chiral chemical potential $\mu_{5}$ (the much debated effect recently, a lot of studies on this contradict each other) has been established in terms of just duality notion, hence reinforce confidence in this result and put it on the considerably more solid ground. 

\end{abstract}

\section{Introduction}

The QCD phase diagram is one of the open questions of modern elementary particle physics that yet to be answered.
The entire QCD phase diagram could not be described currently in the framework of a unified approach. Lattice calculations are very useful for description of the region of zero density and high temperature. However, the so-called sign problem still presents insurmountable difficulties for lattice calculations in the nonzero density region. On the other hand effective theories do not have fundamental background%and as a result don't share the main prominent features with QCD such as a gauge invariance, renormalizability, etc
. Nevertheless, at this moment, effective models are one of the best tool for investigating dense quark matter. At this time one of the most widely used effective model is the Nambu--Jona-Lazinio (NJL) model. 

In addition to temperature and baryon chemical potential $\mu_B$, there are other quantities that describe real quark matter.
It is rather obvious that dense baryonic matter in compact stars has an isospin asymmetry, i.e. where the densities of up- and down quarks are different (it is characterized by isospin chemical potential $\mu_I$). In experiments on heavy-ion collisions, we also can have matter that has an evident isospin asymmetry because of different neutron and proton contents of colliding ions. Using the chiral perturbation theory, it was shown that there is a phase transition at $\mu_I^c=m_\pi\approx 140 MeV$ to the charged pion condensation (PC) phase \cite{Son}. This result was ultimately proved in lattice simulations \cite{BrandtEndrodi}.
There is another imbalance that can matter in different scenarios, it is the chiral imbalance (different densities of right-handed $n_R$ and left-handed $n_L$ quarks). It is expected to appear in heavy-ion collisions on event-by-event basis \cite{Kharzeev:2007jp}. In addition, media with chiral imbalance (chiral media) can be created in various physical systems (in Dirac and Weyl semimetals, in Early Universe, in neutron stars and supernovae) and it is important to study their properties.

Previously, chiral imbalance in the form of chiral isospin ($\mu_{I5}$) and chiral ($\mu_{5}$) chemical potentials were considered in the framework of NJL type model and other effective models \cite{increase,decrease,kkz, Khunjua:2018sro}. In particular, it was shown in Refs \cite{kkz} that in the large-$N_c$ limit ($N_c$ is the number of colors) there is a duality between chiral symmetry breaking (CSB) and charged PC phenomena. It means that the phase portrait of the model under consideration obeys a symmetry with respect to simultaneous transformations, CSB$\leftrightarrow$charged PC and $\mu_{I}\leftrightarrow\mu_{I5}$.

In this paper we explore phase structure of hot quark matter with only chiral or isospin asymmetry at the physical point. In particular the dependence of the (pseudo-)critical temperature, which characterizes the cross-over region of the phase diagram, on the chiral isospin chemical potential $\mu_{I5}$,  $\mu_{5}$ in the framework of the NJL$_4$ model at $m_0\ne 0$. In this case at zero baryon chemical potential, $\mu_B=0$, there is no sign-problem and we have solid results from lattice simulations \cite{Braguta}. The comparison with LQCD results is performed and we discuss the possible use of duality in order to get or reinforce some features of the phase diagram.

\section{The model and its thermodynamic potential}

\subsection{Lagrangian and symmetries}

It is well known that in the framework of effective four-fermion field theories dense and isotopically and chirally asymmetric quark matter, composed of $u$ and $d$ quarks, can be described by the following (3+1)-dimensional NJL Lagrangian
\begin{eqnarray}
 \bar L=\bar q\Big [\gamma^\nu\mathrm{i}\partial_\nu -m_0
+\frac{\mu_B}{3}\gamma^0+\frac{\mu_I}2 \tau_3\gamma^0+\frac{\mu_{I5}}2 \tau_3\gamma^0\gamma^5\Big ]q+ \frac
{G}{N_c}\Big [(\bar qq)^2+(\bar q\mathrm{i}\gamma^5\vec\tau q)^2 \Big
].
\end{eqnarray}
Here $q$ is a flavor doublet, $q=(q_u,q_d)^T$, where $q_u$ and $q_d$ are four-component Dirac spinors as well as color $N_c$-plets of the $u$ and $d$ quark fields, respectively; $\tau_k$ ($k=1,2,3$) are Pauli matrices; $m_0$ is the bare quark mass (for simplicity, we assume that $u$ and $d$ quarks have the same mass); $\mu_B$, $\mu_I$ and $\mu_{I5}$ are chemical potentials which are
introduced in order to study quark matter with nonzero baryon, and isospin and chiral (more specifically chiral isotopic) densities, respectively.

The quantities $\hat n_B\equiv\bar q\gamma^0q/3$, $\hat n_I\equiv\bar q\gamma^0\tau^3 q/2$ and $\hat n_{I5}\equiv\bar q\gamma^0\gamma^5\tau^3 q/2$ are the density operators of the baryon, isospin and chiral isospin charges of the system (1) that corresponds to the  chemical potentials  $\mu_B$, $\mu_I$ and $\mu_{I5}$, respectively. Introducing the particle density operators for $u$ and $d$ quarks, $\hat n_u\equiv q_u\gamma^0q_u$ and $\hat n_d\equiv q_d\gamma^0q_d$, we have
\begin{eqnarray}
\hat n_B=\frac 13\left (\hat n_u+\hat n_d\right ),~~\hat n_I=\frac 12\left (\hat n_u-\hat n_d\right ),~~ \hat n_{I5}=\frac 12\left (\hat n_{uR}-\hat n_{uL}-\hat n_{dR}+\hat n_{dL}\right )=\frac 12\left (\hat n_{u5}-\hat n_{d5}\right ).
\label{2003}
\end{eqnarray}
 In order to find the TDP of the model, we start from a semibosonized version of the Lagrangian (1), which contains composite bosonic fields $\sigma (x)$ and $\pi_a (x)$:
\begin{eqnarray}
{\cal L}\ds =\bar q\Big [\gamma^\rho\mathrm{i}\partial_\rho - m_0 +\mu\gamma^0
+ \nu\tau_3\gamma^0+\nu_{5}\tau_3\gamma^0\gamma^5-\sigma
-\mathrm{i}\gamma^5\pi_a\tau_a\Big ]q
 -\frac{N_c}{4G}\Big [\sigma\sigma+\pi_a\pi_a\Big ].
\label{2}
\end{eqnarray}
Here, $a=1,2,3$ and also we introduced the notations $\mu\equiv\mu_B/3$, $\nu\equiv\mu_I/2$ and $\nu_{5}\equiv\mu_{I5}/2$. From the auxiliary Lagrangian (\ref{2}) one gets the equations for the bosonic fields:
\begin{eqnarray}
\sigma(x)=-2\frac G{N_c}(\bar qq);~~~\pi_a (x)=-2\frac G{N_c}(\bar q
\mathrm{i}\gamma^5\tau_a q).
\label{200}
\end{eqnarray}
Note that the composite bosonic field $\pi_3 (x)$ can be identified with the physical $\pi^0(x)$-meson field, whereas the physical $\pi^\pm (x)$-meson fields are the following combinations of the composite fields, $\pi^\pm (x)=(\pi_1 (x)\mp i\pi_2 (x))/\sqrt{2}$.
Obviously, the semibosonized Lagrangian ${\cal L}$ is equivalent to the initial Lagrangian (1) when using the equations (\ref{200}).

In the present paper we suppose that in the ground state of the system the quantities $\vev{\sigma(x)}$ and $\vev{\pi_a(x)}$ do not depend on spacetime coordinates $x$ and we assume the following ansatz
\begin{eqnarray}
\vev{\sigma(x)}=M-m_0,~~~\vev{\pi_1(x)}=\Delta,~~~\vev{\pi_2(x)}=0,~~~ \vev{\pi_3(x)}=0. \label{06}
\end{eqnarray}

 One can obtain the following expression for the TDP $\Omega_T(M,\Delta)$
\begin{equation}
\Omega_T (M,\Delta)
=\Omega (M,\Delta)
-T\sum_{i=1}^{4}\int_{0}^{\Lambda}\frac{p^2dp}{2\pi^2}\Big\{\ln(1+e^{-\frac{1}{T}(|\eta_{i}-\mu|)})+\ln(1+e^{-\frac{1}{T}(|\eta_{i}+\mu|)})\Big\},\label{260}
\end{equation}
%where 
%$$
%\text{where}\,\,\,\,\, \Omega (M,\Delta)
%=\frac{(M-m_0)^2+\Delta^2}{4G}-\frac{1}{2\pi^2}\sum_{i=1}^{4}\int_{0}^{\Lambda}p^2\big (|\eta_{i}|+\theta(\mu-|\eta_{i}|)(\mu-|\eta_{i}|)\big )dp.\label{26}
%$$
where $\Omega (M,\Delta)$ are the TDP in case of zero temperature and $\eta_{i}$ can be found analytically, we will not include the procedure here, they can be found in \cite{Khunjua:2018sro}.

One can show that in the chiral limit, the TDP is invariant with respect to the so-called duality transformation:
\begin{equation}
{\cal D}:~~~~M\longleftrightarrow \Delta,~~\nu\longleftrightarrow\nu_5.~~~~~~~~16
 \label{16}
\end{equation}
 We use the following, widely used parameters:
\begin{eqnarray}
m_0 = 5,5 \,{\rm MeV};\qquad G=15.03\, {\rm GeV}^{-2};\qquad \Lambda=0.65\, {\rm GeV}.\label{fit}
\end{eqnarray}

\section{Phase structure in NJL model and lattice QCD}
Let us first consider the case of only one non-zero isospin chemical potential  $\mu_I\equiv 2\nu$. Let us plot the $(\nu,T)$-phase diagrams of the model at $\mu=0$ in order to compare our results with lattice investigations, in case there is no sign problem. One can see this diagram at Fig. 1. One can see that at zero temperature and small isospin chemical potential the system shows the so-called
Silver Blaze phenomenon (the ground state of the system is not affected by increase of chemical potential). When the
value of isospin chemical potential crosses the threshold, $\mu_I=\frac{m_{\pi}}{2}$, charged pions can be created and pion condensation appears. The associated phase transition is expected to be of second order in the O(2)
universality class \cite{Son}. At the Fig. 2 the same phase diagram obtained in LQCD is depicted and one can see that the NJL model results are in agreement with first principle lattice calculations \cite{BrandtEndrodi}.

\begin{figure}%[h!]
\includegraphics[width=0.535\textwidth]{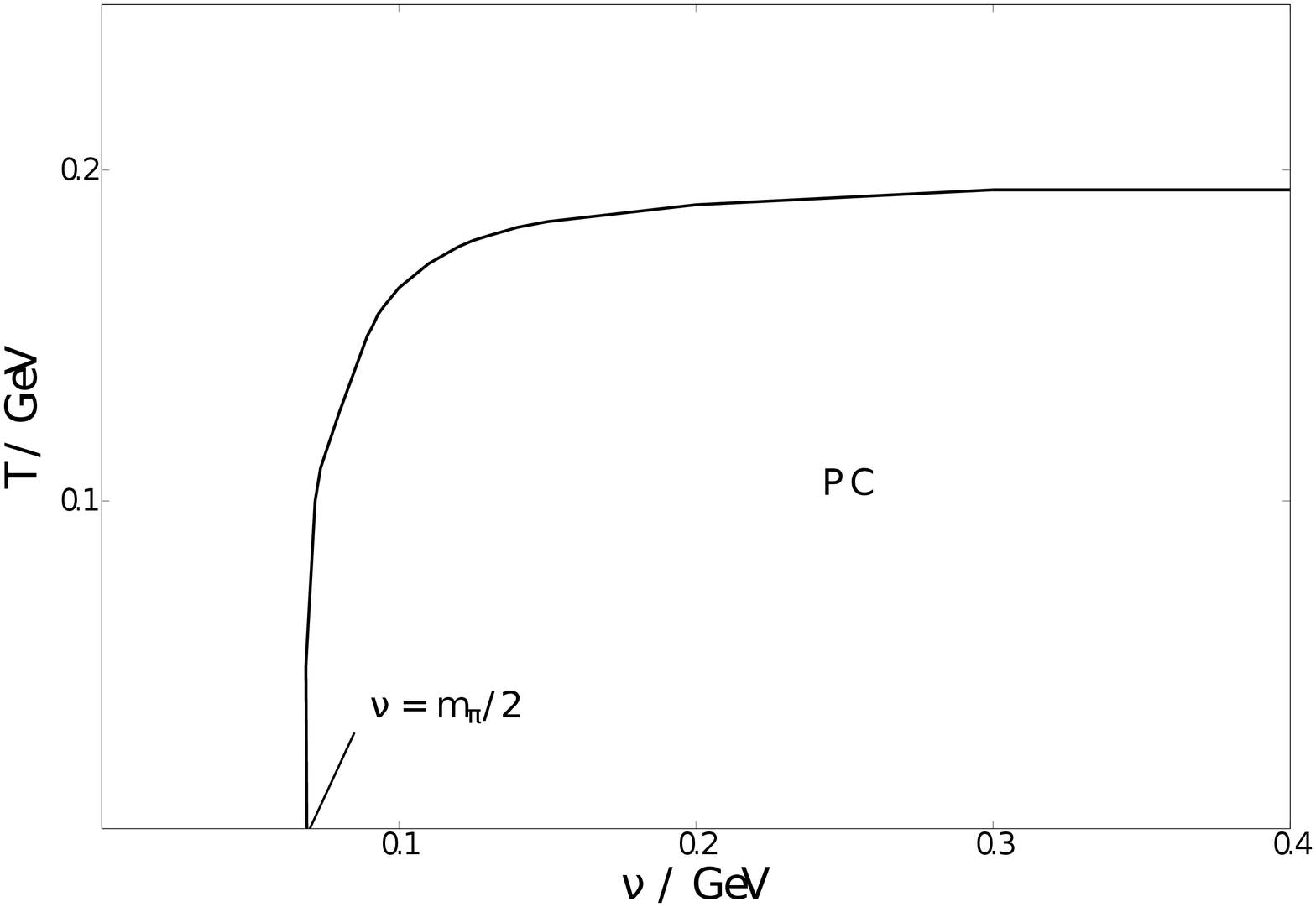}
 \hfill
\includegraphics[width=0.58\textwidth]{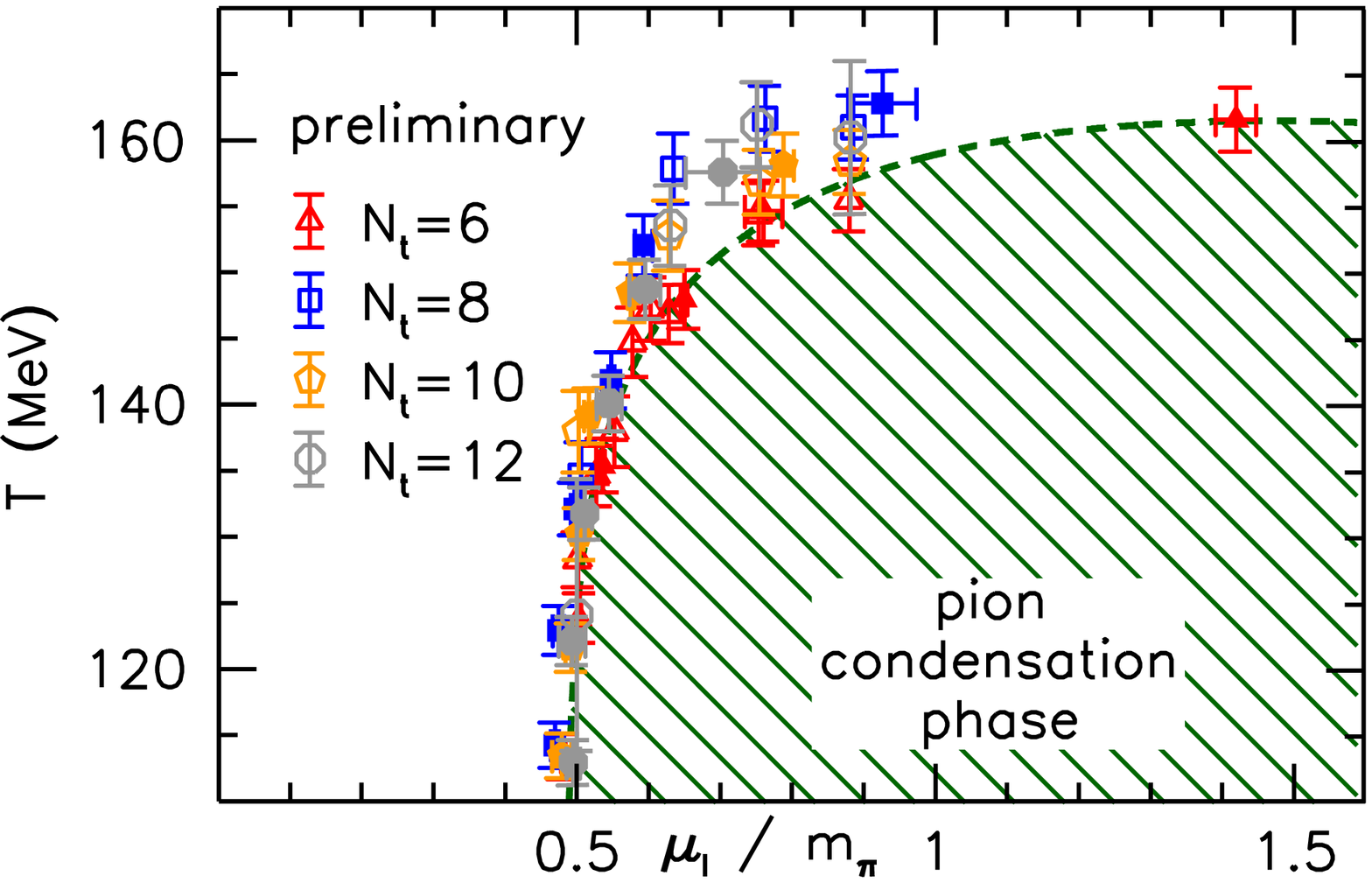}\\
\label{fig1}
\parbox[t]{0.4\textwidth}
{\caption{ The ($\nu,T$)-phase portraits at $\mu=\nu_5=0\,{\rm MeV}$. Results obtained in NJL model.} }\hfill
\parbox[t]{0.4\textwidth}{
\caption{ The ($\nu,T$)-phase portraits at $\mu=\nu_5=0\,{\rm MeV}$. Results obtained in LQCD. Taken from PoS LATTICE 2016 (2016) 039,  Phys. Rev. D 97, 054514 (2018).
}}
\end{figure}

\begin{figure}
\includegraphics[width=0.57\textwidth]{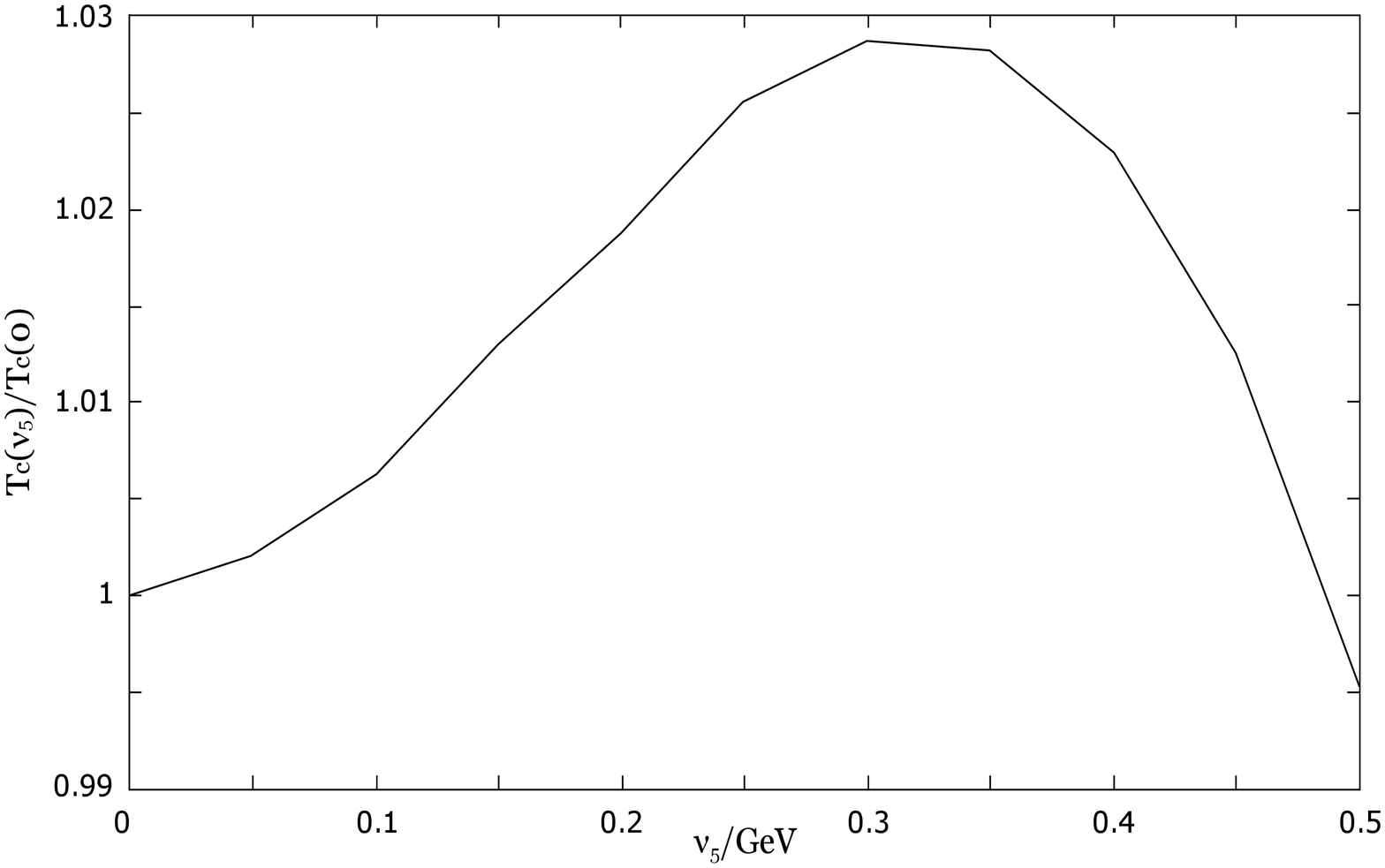}
 \hfill
\includegraphics[width=0.56\textwidth]{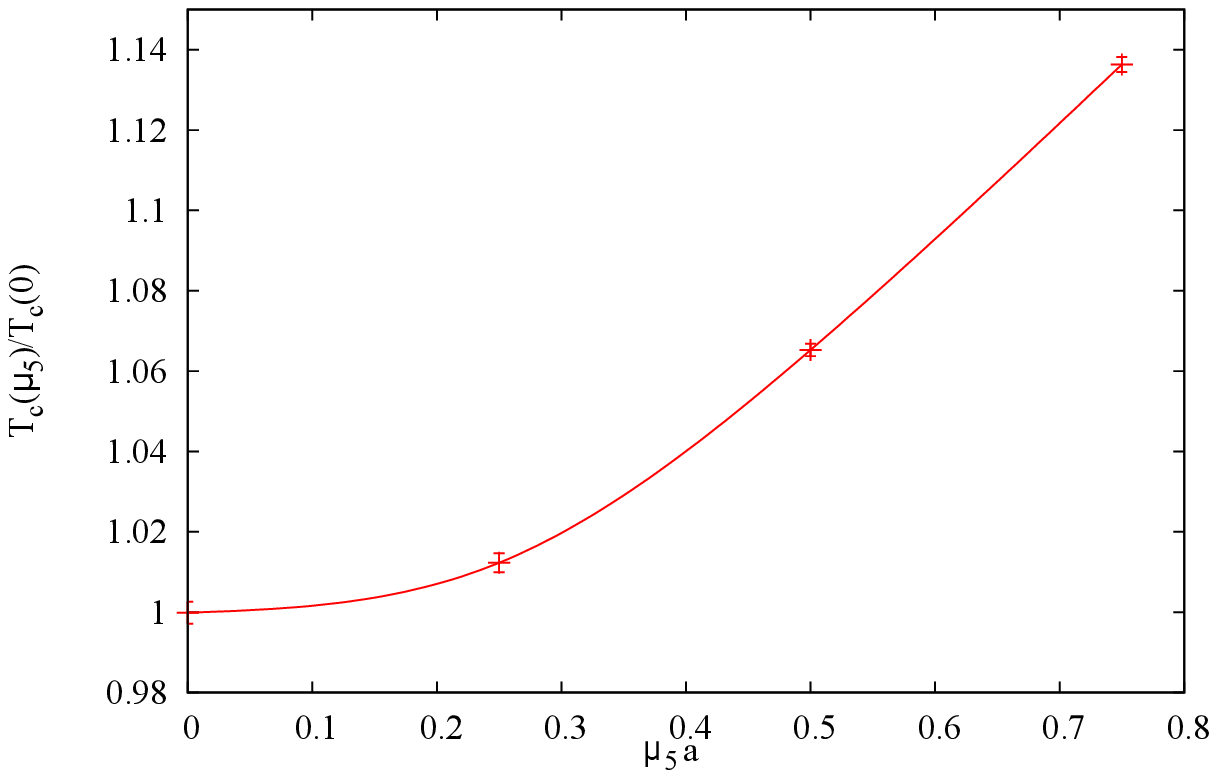}\\
\label{fig1}
\parbox[t]{0.4\textwidth}
{\caption{ The ($\nu_5(\mu_{5}),T$)-phase portraits at $\mu=\nu=0\,{\rm MeV}$. Results obtained in NJL model.} }\hfill
\parbox[t]{0.4\textwidth}{
\caption{ The ($\nu_5(\mu_{5}),T$)-phase portraits at $\mu=\nu=0\,{\rm MeV}$. Results obtained in LQCD. Taken from Phys.\ Rev.\ D {\bf 93}, 034509 (2016).
}}
\end{figure}

%Up to now we talked about the consideration of the critical temperature $T^{PC}_c$ (at $\mu=0$ and different fixed values of $\nu$) of the second-order phase transition from the charged PC to the ApprSYM phase at $\nu_5=0$. So we considered the section $\nu_5=0$, $\mu=0$ of the full phase diagram $(\mu, \nu, \nu_5)$ which can be considered in lattice QCD simulations. 
Now let us consider the dual section $\nu=0$, $\mu=0$ of the full phase diagram $(\mu, \nu, \nu_5)$, more specifically %let us consider the section $\nu=0$, $\mu=0$. 
the behavior (at $\mu=0$ and $\nu=0$) of the pseudo-critical temperature $T_c(\nu_{5})$. This transition is usually called a cross-over because different physical parameters, such as the dynamical quark mass $M_0$ etc, of the CSB phase smoothly (without jumps), but rather sharply go over to the corresponding parameters of the ApprSYM phase. Therefore, in this region, there occurs not a true phase transition with corresponding critical temperature, etc., but rather a {\it pseudo}-phase transition characterized by a {\it pseudo}-critical temperature $T_c(\nu_5)$, etc. The case of only non-zero chiral chemical potential  $\mu_{5}$ can be considered on the lattice because in this case there is no sign problem.
We plan to compare our NJL model results with lattice simulations of QCD phase diagram with chiral $\mu_{5}$ chemical potential but we considered the NJL model at non-zero chiral isospin $\mu_{I5}$ chemical potentials. In the most general case, the chiral asymmetry of dense quark matter is described by two chemical potentials, chiral $\mu_5$ and chiral isospin $\mu_{I5}\equiv 2\nu_5$ chemical potential. So how can we compare different results and even draw some conclusions. The point is that from the recent paper \cite{Khunjua:2018sro}, where the phase structure of this model was investigated in the chiral limit in a more general case with all four nonzero chemical potentials $\mu, \mu_5, \mu_I, \mu_{I5}$, we know that  ($\mu_5$, $T$) and ($\nu_5$, $T$) phase diagrams contain only CSB phase, and it has been established in \cite{Khunjua:2018sro} that in addition to the dual symmetry (\ref{16}) the TDP of the NJL$_4$ model is invariant with respect to a transformation ${\cal D}_M:~\mu_5\longleftrightarrow\nu_5\,,~~ \Delta=0$. This duality states that  $\mu_{5}$ chemical potential and chiral isospin $\mu_{I5}$ chemical potential %and  chiral $\mu_{5}$ chemical potential
influence chiral symmetry breaking phenomenon in exactly the same way even in the physical point. So the phase diagrams  ($\mu_5$, $T$)  and   ($\nu_{5}$, $T$)  are exactly the same due to this duality and this enable us to compare our results with lattice QCD ones.

The plot of the  {\it pseudo}-critical temperature $ T_c(\nu_{5})\big(= T_c(\mu_{5})\big)$ is depicted at Fig. 3, the lattice QCD result for $T_c(\mu_{5})$ is shown at Fig. 4. It is easy to see from the plot of Fig. 3, that the pseudo-critical temperature of the NJL model increases for $\mu_5<\mu^\ast_5 \lessapprox 350$ MeV. Above this value it drops down, but at $\mu_5>\mu^\ast_5$, in our opinion, the NJL$_4$ model does not provide very reliable predictions, because $\mu_5$ is near the cutoff $\Lambda$. One can see that up to the values of $\mu_5=300$ MeV there is a good qualitative and even quantitative agreement with the lattice results (see Fig.4).
 
So let us gaze at all this from the general picture viewpoint. We have two lattice simulation results,  ($\mu_I$, $T$) and  ($\mu_5$, $T$) phase diagrams  (Fig.2 and Fig 4. respectively). These phase diagrams have been also obtained in the NJL model and the results are in good agreement with lattice QCD simulations. But in terms of NJL model we can consider the general case and we know that there is the duality between CSB and charged PC phenomena in the leading order of the large-$N_{c}$ approximation. So the particular phase diagrams  ($\mu_I$, $T$) and  ($\mu_5$, $T$) should be dual to each other, but the duality in the case of the physical point is only approximate, although, it is valid with a good precision \cite{Khunjua:2018jmn}. Since the particular phase diagrams  ($\mu_I$, $T$) and  ($\mu_5$, $T$) in these two approaches agrees, one can conclude that the duality can be observed in the lattice QCD simulations.
And this put the notion of the duality on another level of confidence, for it is observed in terms of the toy (1+1) dimensional NJL model, effective (3+1) dimensional NJL model, lattice QCD simulations and similar dualities has been observed in the large $N_{c}$ orbifold equivalences approach \cite{Hanada:2011ju, Hanada:2011jb}.
Comparison to the lattice QCD is important not only due to the fact that it is ab initio method for dealing with QCD but because it does not make use of, for example, large $N_{c}$ approximation (as in NJL models or in large $N_{c}$ orbifold equivalences approaches).

The question of catalysis of chiral symmetry breaking by chiral chemical potential (increase of {\it pseudo}-critical temperature) is a rather debated one and there are a number of papers that predicted that critical temperature decrease with increase of chiral chemical potential \cite{decrease} as well there are a number of papers that supports the catalysis \cite{increase}. Lattice QCD results is probably more trustworthy and it predicts catalysis. But lattice simulations are performed with unphysically large value of pion mass so it does not give ultimate result yet as well. But the catalysis of chiral symmetry breaking by chiral chemical potential can be established in terms of duality notion, let us elaborate on that. As we have talked about the  ($\mu_I$, $T$)  phase diagram is well established one and the duality fails only in the region of small isospin and chiral chemical potentials (smaller than half of the pion mass), but works quite well for the larger values. But at the ($\mu_I$, $T$)  phase diagram in the region of isospin chemical potential larger than half of the pion mass the critical temperature increases when $\mu_{I }$ is raised and the duality here is a good approximation, so the critical temperature at the duality conjugated  ($\mu_5$, $T$)  phase diagram should increase with rising of $\mu_{5}$ as well.

\section{Summary}
Let us summarize the core results of our paper.

\begin{itemize}

\item The particular cross sections of the full phase portrait $(\mu, \nu, \nu_5)$ are in qualitative accordance with the recent lattice simulations \cite{BrandtEndrodi, Braguta}. So it has been established that lattice QCD results support the existence of the approximate duality in real QCD.

\item The rise of pseudo-critical temperature with increase of chiral chemical potential $\mu_{5}$ has been established in terms of duality notion and well explored the results of lattice QCD and different approaches on phase structure of isotopicaly imbalanced quark matter.  It gives additional argument in favour of this behaviour of the pseudo-critical temperature and it is of importance because, although the lattice results confirming this behaviour are conclusive, the pion mass that is used in these simulations is still quite high and well above the physical pion mass and our results are made at the physical point with the right value of the pion mass. 
\end{itemize}

\section{Acknowledgements}
R.N.Z. is grateful for support of the Foundation for the Advancement of Theoretical Physics and Mathematics
BASIS grant and Russian Science Foundation under the grant No 19-72-00077

\end{document}